# Mapping of SOA and RUP: DOA as Case Study

Shahid Hussain
Namal University, Mianwali

Sheikh Muhammad Saqib
IC IT, Gomal University

Dr. Bashir Ahmad
ICIT, Gomal University

Dr. Shakeel Ahmad
ICIT, Gomal University

**Abstract-** SOA (Service Oriented Architecture) is a new trend towards increasing the profit margins in an organization due to incorporating business services to business practices. Rational Unified Process (RUP) is a unified method planning form for large business applications that provides a language for describing method content and processes. The well defined mapping of SOA and RUP leads to successful completion of RUP software projects to provide services to their users. DOA (Digital Office Assistant) is a multi user SOA type application that provides appropriate viewer for each user to assist him through services. In this paper authors proposed the mapping strategy of SOA with RUP by considering DOA as case study.

## 1. INTRODUCTION

SOA is an environment for dynamic invention and use of services over connected nodes in a network. SOA discipline may be used to build infrastructures finding needs and those with capabilities through services across network. A decision point for any policies and contracts that may be in force can be done in SOA [1][2]. An enterprise-scale IT architecture is called SOA which is used for linking resources on demand. These resources are made available to participants in a value net.[3]. RUP is a consistent methodology that supports steady development and focuses on large business projects and provides a collection of customizable techniques and practices for developing object oriented solution[4]. If proper adjustment on the design side of methodology is made, RUP can provide greater chance for SOA type application [4]. In this paper, authors give the concept of development by mapping practices of SOMA framework and RUP phases.

## 2. RATIONAL UNIFIED PROCESS (RUP)

RUP is a software engineering process model, which provides a disciplined approach to assigning tasks and responsibilities within a development organization. The goal of RUP is to produce high quality software that meets the needs of its end users within a predictable schedule and budget [5]. There are two structure approaches for RUP: Static Structure, Dynamic Structure. Static structure of RUP comprise on four elements which are: Role, Activities, Artifacts, Disciplines/workflows. Dynamic structure of RUP comprises of phases and iterations in each phase: Inception, Elaboration, Construction, Transition.

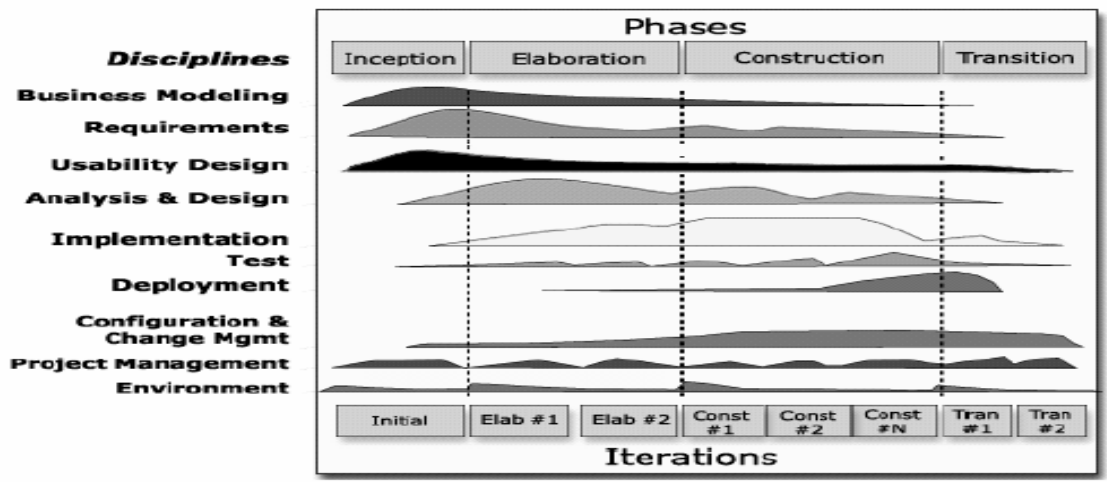

Fig-1. RUP Activities Diagram Adopted from [5]

## 3. SOMA'S FRAMEWORK FOR SOA

SOMA's framework for SOA development consists of four activities such as Service Identification, Service Specification, Service Realization and Service Deployment as shown in Fig-2. Service Identification (SI), point to identification and selection of candidate services; Service Specification (SS) refers to Specification of the set of services by developing a Service Model; Service Realization (SR), refers to the designing of service component in design model and finally Service Deployment(SD refers to the transferring of the services to the production environment.

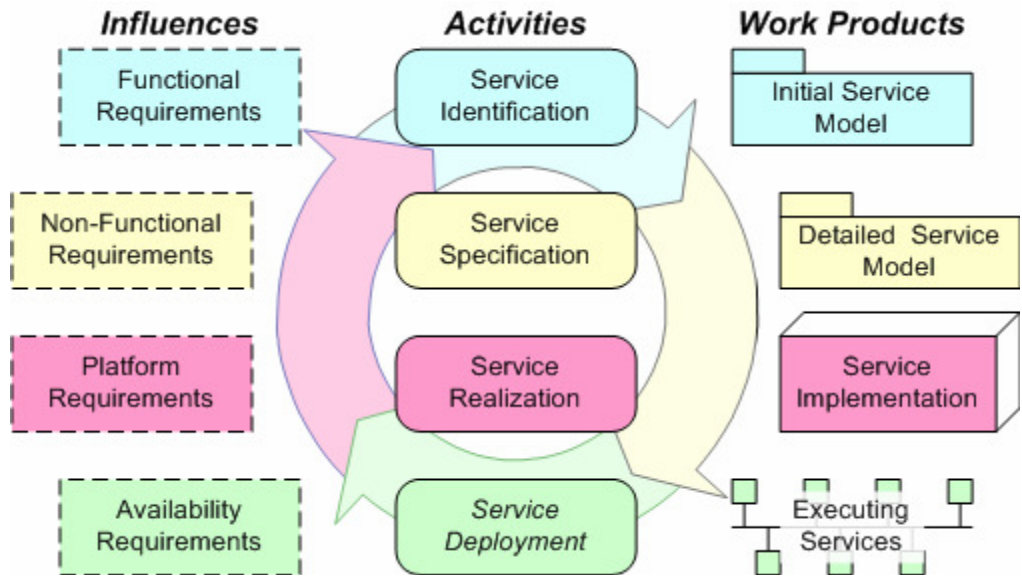

**Fig-2 The SOMA framework:** Adopted from [6]

Every RUP project follows a strictly linear scheme of four phases: Inception, Elaboration, Construction and Transition. Each of these phases, except for inception, encompasses all the activities service identification, service specification and service realization. During the inception phase, emphasis is given to determine the scope of project not in term of services. In normal development of RUP project, inception is considered as core phase to understand the whole concept but in SOA methodology inception can be namely as a part of the Service realization activity [6].

## 4. MAPPING STRATEGY : DOA AS CASE STUDY

DOA is a SOA based application developed in .Net framework which can be used in any office and can assist all employees. Each user can perform different activities of DOA after authentication. The list of services which are provided by DOA are classified and shown in Fig-3.

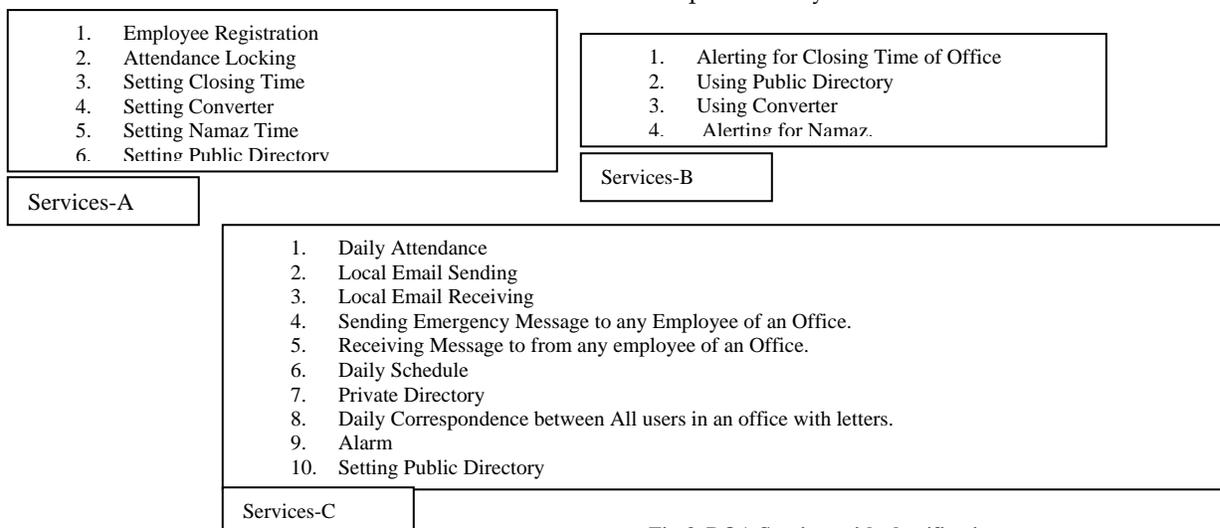

**Fig-3. DOA Services with classification**

The first category A comprises on services which are used by administrator and second category B comprises on services which are used by all users. Similarly category C's services used by user privately.

TABLE-1. COMPARISON OF RUP PHASES AND SOA ACTIVITIES.

| Recommended Phases of RUP | SOA Activities | Description | RUP / SOA | Discipline |
|---|---|---|---|---|
| Inception | Nill | Determining scope of project. | | Nill |
| | | After scope of project, no attention is made on Service. | | |
| Elaboration | Service Identification Service Specification | Lifecycle of Architecture miles stone | | Business Modeling, Requirement, Analysis & design, Implementation, Project management |
| | | Selection of Candidate Services and model with different operations, which is building blocks for architecture. | | |
| Construction | Service Realization | Initial Operational capability milestone | | Requirement, Analysis & design, Project management, Configuration and Change Management |
| | | Designing Service | | |
| Transition | Service Deployment | Product Release Miles Stone | | Deployment, Project management, Configuration and Change Management |
| | | Service plug-in to System | | |

Administrator of DOA can register any new employee by assigning user ID and Password. The new employee can login to DOA application from any PC and then can set his daily schedule and private directory etc. Similarly new user can send local email or emergency message or any letter (diary and dispatch will be dine automatically) to any other employee by using their ID. Each employee can make daily attendance on DOA over their PC in specified time which is set by administrator. After specified time, attendance will be locked by administrator and each employee would be unable for attendance. Table-1 shows the comparison of SOA and RUP activities.

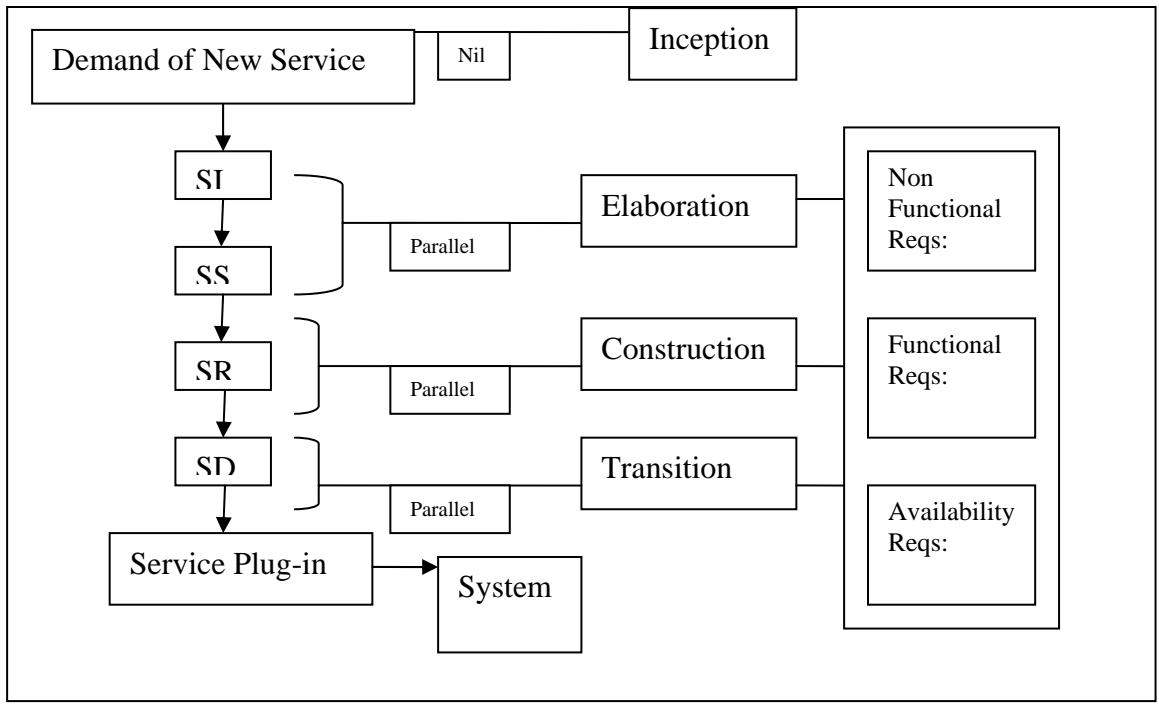

**Fig-4. Mapping of SOA and RUP Activities**

After studying RUP and SOMA's frame work, authors observe that RUP phases can involved many activities work of SOAM's framework. Fig-4. shows the mapping process of RUP and SOA. During elaboration phase of RUP, services of DOA are identified which are shown in Table-2.

TABLE-2 IDENTIFICATION OF DOA SERVICES.

| Services | Influence | Functional Req: / Non Functional Req: | Couple Service |
|---|---|---|---|
| Services A | Appropriate Viewer | | Depends on Administrator Activities |
| | Product Requirement, Organizational Requirement, External Requirement | | |
| Services B | Appropriate Viewer | | B1 -> A3<br>B2 -> A6, B10<br>B3 -> A4<br>B4 -> A5 |
| | Product Requirement, Organizational Requirement | | |
| Services C | Appropriate Viewer | | C1 to C9 -> A-1 |
| | Product Requirement, Organizational Requirement, External Requirement | | |

The service model is the outcome of SOA activities which shows the specification of service by defining input, output and relevant operations. The service model for specification of DOA services is shown in Table-3. There are many services of DOA application but authors are showing the specification of only local mail service and daily schedule service. The letter I, D and M in operation column of Table-3 represent the Insertion, Deletion and Modification operations for a specific service.

TABLE-3. SERVICE MODEL FOR DOA

| Services | In Put. | | Out Put | Operation |
|---|---|---|---|---|
| Email Sending | TUID | ID for a user to whom email is sending | Any user can check their emails through appropriate interface. | I,D |
| | FUID | ID for a user who is sending email | | |
| | SUB | Subject about email | | |
| | MSG | Detail of Message | | |
| | DT | Date and Time of Sending email | | |
| Daily Schedule | UID | ID for a user who is login with DOA and Setting his schedule | A user who set his daily schedule will be alert automatically after login to DOA with his ID and password. | I,D,M |
| | DT | Setting Date for alert | | |
| | TM | Setting Time for alert | | |
| | Reminder | A message for alert on time of particular date | | |
| | Status | Alerting ON or OFF | | |

The Table-4 shows the realization of targeted services such Local Email and Daily Schedule of DOA.

TABLE-4. REALIZATION OF DOA SERVICES

| Services | Service Component | |
|---|---|---|
| Local Email | Sending Email | 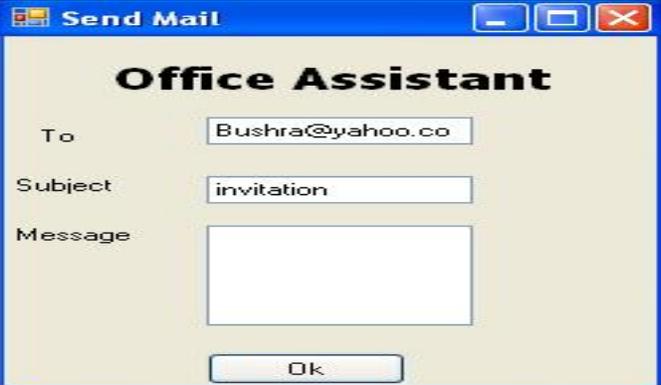 |
| | Receiving Email | 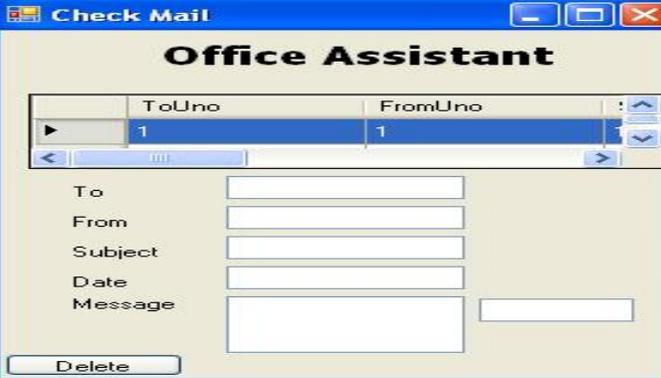 |
| Daily Schedule | | 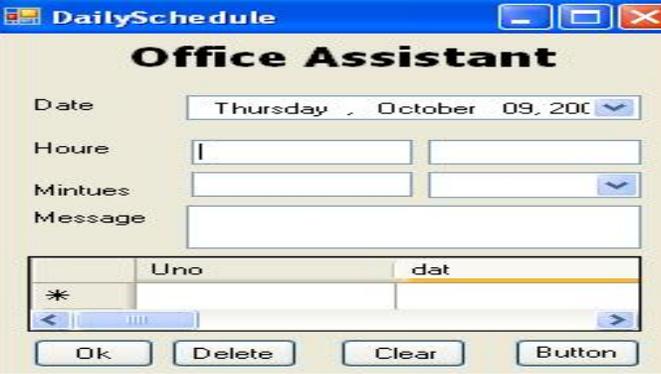 |

Finally the targeted services or any new service of DOA are deployed through SOA model integrated with RUP shown in Fig-5.

## 5. CONCLUSION

RUP process model uses the unified process and object oriented methodology to develop an enterprise level applications. Now a days many organizations focus on SOA application due to its benefits of proper design and development strategy. In this paper authors proposed a mapping strategy for SOA and RUP and developed a student level project DOA (example of SOA application) and by defining mapping strategy.

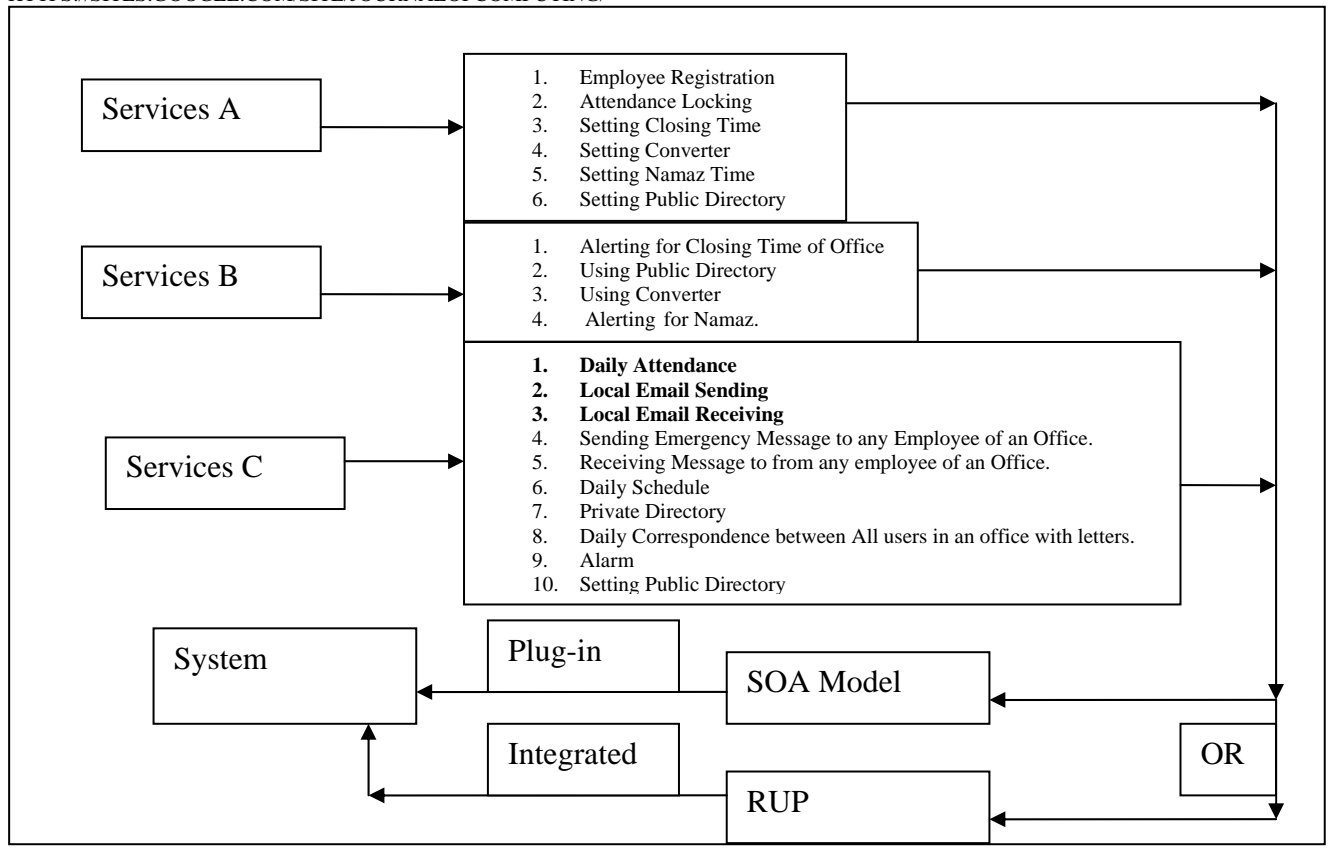

**Fig-5. DOAs' Service Deployment**